Supplementary Materials for

# How to Avoid Reidentification with Proper Anonymization

## *Comment on "Unique in the shopping mall: on the reidentifiability of credit card metadata"*

**Authors:** David Sánchez*, Sergio Martínez, Josep Domingo-Ferrer

**Affiliations:**

UNESCO Chair in Data Privacy, Department of Computer Engineering and Mathematics, Universitat Rovira i Virgili (URV), Av. Països Catalans, 26, E-43007, Tarragona, Catalonia.

*Correspondence to: E-mail: david.sanchez@urv.cat.

**This document includes:**
Background and Motivation
Materials and Methods
Figures S1-S7
Tables S1-S10
Algorithms S1-S4
Summary
References

**Other Supplementary Material for this manuscript includes the following:** (available at http://crises-deim.urv.cat/opendata/SPD_Science.zip):
Datasets for replication of all the experiments and results (text files);
Open source code (java) of the described algorithms, with usage examples.

**Acknowledgments:**
Thanks go to Jane Bambauer, Ann Cavoukian, Khaled El Emam, Krish Muralidhar and Vicenç Torra for useful reviews and discussions.
The following funding sources and grants are gratefully acknowledged: European Commission (H2020-644024 "CLARUS"), Spanish Government (TIN2012-32757 and TIN2014-57364-C2-1-R), Government of Catalonia (2014 SGR 537 and ICREA-Acadèmia award to J. Domingo-Ferrer), and Templeton World Charity Foundation (TWCF0095/AB60). The opinions expressed in this paper are the authors' own and do not necessarily reflect the views of any funder or UNESCO.

## Supplementary Materials

### Background and Motivation.

The availability of data on individuals (such as electronic healthcare records, census information, etc.) is of utmost importance in many areas of research, especially in fundamentally empirical disciplines studying humans and their behavior (health sciences, social sciences, etc.). Because of the inherent sensitivity of many of these data, they should be appropriately transformed before release in order to protect the privacy of the persons to whom they correspond (data subjects). The main objective should be to prevent anyone from reidentifying a subject from her released record, because this would disclose the subject's confidential information (salaries, diagnoses, etc.). This process, known as data de-identification or anonymization, has been challenged by recent studies (*1, 2*), which claim that knowing a small amount of personal information on an individual is enough to reidentify her.

Specifically, in the study by de Montjoye et al. (*1*) (dM in the sequel), the authors concluded that, for most customers in a de-identified credit card transaction database, knowing the shop (location), date and the price of four purchases by the customer was enough to reidentify her. Reidentification risk was measured according to "unicity" (*2*) (a neologism for the plain old uniqueness notion) which, given a number of personal features assumed known to an external entity (an attacker), counts the number of individuals having a unique combination of these features. Because of the high "unicity" observed in a sample of the de-identified dataset used in their study, dM implemented some additional anonymization strategies to reduce the detail of data (such as clustering shops) which fell short of sufficiently reducing "unicity". From this, dM drew worrisome conclusions about the effectiveness of anonymization methods for data release, like "*This means that these [de-identified financial] data can probably be relatively easily reidentified if released in a simple anonymized form and that they can probably not be anonymized by simply coarsening of the data*". Such conclusions have triggered a plethora of bold media statements such as "*relying on anonymization algorithms to scrub our personal information [...] is not currently a viable solution*" (*3*), "*the old model of anonymity doesn't seem to be the right model when we are talking about large-scale metadata*" (*4*), "*the open sharing of [de-identified metadata] datasets is not the future*" (*5*) "*Credit card study blows holes in anonymity*" (*6*), etc. In view of the alleged lack of effective anonymization solutions, dM also highlighted the need to develop "*privacy-conscientious technologies, as well as the need for more research in computational privacy*" and "*to reform our data protection mechanisms beyond anonymity and toward a more quantitative assessment of the likelihood of reidentification*".

Data owners (and subjects!) may be dissuaded by the above statements from releasing any data on individuals (also known as microdata), which would be catastrophic for many research efforts (*7*), let alone the open data initiatives promoted by administrations for the sake of transparency and economic growth (e.g., the U.S. Open Government Initiative). Moreover, the general public may become persuaded that anonymization cannot cope with protecting data releases.

As long-standing researchers on data privacy, we strongly believe that many of the conclusions of dM are unfounded and derive from a loose reinvention of well-known privacy concepts (e.g., "unicity" is a reinvention of uniqueness), a misconception of the reidentification attack, a severely flawed implementation of anonymization, and a general unawareness of the relevant literature in data anonymization. Our aim in this study is to bring clarification and restore trust in data anonymization by showing how to properly anonymize data.

## Materials and Methods.

**The dataset.** To conduct our study, we first asked dM to share the raw financial data they used to evaluate their "unicity"-based reidentification risk, which consist of credit card transactions of 1.1 million users during 3 months, detailing the shop, time and price of each transaction. However, as they already warned in their article, the authors confirmed they could not share their dataset for contractual and privacy reasons. This contradicts the journal's data availability policy and makes it impossible to reproduce their claimed successful attack, let alone evaluate it against *sound* anonymization mechanisms. Hence, we were forced to look for a transaction dataset with similar structure and "unicity"/reidentification risk properties.

Given the high sensitivity of health-related information (*8*), and in order to provide transparency and true reproducibility, we chose a publicly available dataset consisting of patient discharge data (PD from now) collected from all Californian hospitals in 2009. This dataset was released by the California Office of Statewide Health Planning and Development (OSHPD) and is available at: http://www.oshpd.ca.gov/HID/DataFlow/index.html. Each record corresponds to a patient discharge and is comma-delimited into attributes describing personal and clinical features of the patient. The data provided by the OSHPD is divided into three separate files based on the geographic location of the hospital: Los Angeles County (containing 1,183,718 records), Southern California (including Imperial, Orange, Riverside, San Bernardino, San Diego, Santa Barbara and Ventura counties, with a total 1,201,553 records) and Northern California (remaining counties, with 1,599,895 records). In our study, we considered all these three files as a single dataset (which we called the PD dataset).

**Description of the attributes.** The PD dataset consists of 3,985,166 records in total. Each record describes the personal and clinical details of the patients with 38 attributes (see the complete description in (*9*)). From these, there are 9 attributes detailing census and spatiotemporal features of the admission, 12 attributes about internal facility administrative information (such as the expected type of payment or the type of coverage) and the remaining 17 attributes describe the clinical conditions of the patient. In our study we considered the 9 census and spatiotemporal features as quasi-identifying attributes, because they represent the kind of information that could be known and used by an external entity to reidentify the patient. As confidential attributes, we considered the charges of the medical service and the diagnosis related to the cause of admission, which are especially sensitive. We classified quasi-identifying attributes into *census features* (age, sex, ethnicity, race, ZIP code, and county), *spatial features* (hospital ID) and *temporal features* (admission quarter and length of stay). Table S1 describes in detail the considered attributes and Table S2 provides some examples of such attributes.

**Table S1. Description of the 11 patient discharge attributes considered in our study**

| Attribute | Type | Description |
|---|---|---|
| oshpd_id | Spatial | Hospital ID: a unique 6-digit identifier assigned to each hospital |
| age_yrs | Census | Age in years: age of the patient at admission (from 0 to 85); ages greater than 85 years are assigned a value of 85 years |
| Sex | Census | Gender of the patient for the current admission, 1-digit encoded |
| Ethncty | Census | Ethnicity (self-reported) of the patient, 1-digit encoded |
| Race | Census | Patient's racial background (self-reported), 1-digit encoded |
| Patzip | Census | ZIP code: the patient's 5-digit ZIP code of residence |
| Patcnty | Census | County: the patient's county of residence, 2-digit encoded |
| Los | Temporal | Length of stay: total number of days from admission to discharge |
| adm_qtr | Temporal | Admission quarter: the calendar quarter the patient was admitted, 1-digit encoded |
| charge | Confidential | Total charges for services rendered during the length of stay |
| diag_p | Confidential | Diagnosis: Condition established to be the chief cause of the admission of the patient to the hospital for care. Diagnoses are coded according to the ICD-9-CM |

**Table S2. Example of attribute values from the Patient Discharge (PD) dataset**

| *Quasi-identifiers* | | | | | | | | | *Confidential attributes* | |
|---|---|---|---|---|---|---|---|---|---|---|
| *Spatial feature* | *Census features* | | | | | | *Temporal features* | | | |
| **Hospital** | **Age** | **Gender** | **Ethnicity** | **Race** | **Zip code** | **County** | **Admission quarter** | **Length of stay** | **Charges** | **Diagnosis** |
| Cedars-Sinai | 42 | Male | Non-hispanic | White | 93722 | Fresno | 1 | 33 | $505,785 | Pneumonia |
| Cedars-Sinai | 35 | Male | Non-hispanic | White | 93611 | Fresno | 2 | 47 | $1,083,683 | Meningitis |
| Cedars Sinai | 52 | Male | Non-hipanic | Black | 93561 | Kern | 2 | 13 | $252,303 | Necrosis |
| Cedars Sinai | 70 | Female | Hispanic | White | 93560 | Kern | 2 | 1 | $107,179 | Breast Cancer |
| Cedars Sinai | 57 | Male | Non-hispanic | White | 93555 | Kern | 4 | 2 | $34,392 | Kidney Cancer |
| Cedars Sinai | 61 | Male | Non-hispanic | White | 93551 | Los Angeles | 2 | 2 | $24,846 | Stomach Cancer |

**Appropriateness of the dataset.** This dataset is more appropriate than dM's to measure the *true plausible* reidentification risk because:

- dM's dataset probably includes only a *fraction* of the general population, since the whole population of the undisclosed country in which the dataset was compiled may well be larger than 1.1 million. As highlighted in a reply to dM by Barth-Jones et al. (*10*), with a non-exhaustive sample, the uniqueness/unicity of an individual in the sample does not imply uniqueness in the population and hence does not allow *unequivocal* reidentification; assuming otherwise does clearly overestimate the reidentification risk. In contrast, the PD dataset includes *all* the patients of the Californian hospitals; thus, if a person belongs to the population of patients of Californian hospitals during 2009, her clinical record *appears* in the dataset. Then, if that person's features are unique in the dataset *and she can be reidentified*, reidentification is unequivocal.
- dM assume that attackers can gather quite detailed and highly "dynamic" [quasi-identifying] and circumstantial features of the individuals (dates, locations and approximate prices of small transactions in several shops). As also noted in (*10*), these features would actually be quite difficult to compile by an attacker on a target individual, because they are rarely released, which makes reidentification less feasible. On the contrary, our PD dataset includes mostly "static" QIs, more plausibly known by an attacker (such as the set of census features of the patient).

**Masked Values.** The data provided by OSHPD are already protected according to the "safe harbor" rules on medical data protection defined by the Health Insurance Portability and Accountability Act (HIPAA) (*8*). This means that records that originally had unique combinations of certain demographic variables have already been partially masked. The masking process consists in a combination of value coarsening and value suppression (suppressed values are replaced with a wildcard (*)). In order to retain the utility of the published dataset, the smallest possible number of values should be masked. To do so, attributes are masked in the sequence and in the way described by Table S3 (extracted from (*9*)). Specifically, in the first place, the numerical age value is coarsened to one of 20 possible categories (ranges); if doing so does not yet yield a non-unique record, ethnicity is suppressed (replaced with a wildcard *); if the record is still unique, then race is suppressed; if still needed, sex is suppressed; the next step is to further coarsen age to one of 5 possible categories, etc. From Table S3, it can be seen that masking for age and ZIP code means successive coarsening and eventually suppression (ZIP code is first coarsened from 5 to 3 digits). For the other attributes, masking directly means suppression.

**Table S3. Masking order for unique records; attribute values are masked in this order until the record becomes non-unique with regard to these attributes**

| Masking order | Masked attribute |
|---|---|
| 1 | Age in years; age range (among 20 categories) remains |
| 2 | Ethnicity |
| 3 | Race |
| 4 | Sex |
| 5 | Age range (20 categories); age range (among 5 categories) remains |
| 6 | Age range (5 categories) |
| 7 | Admission quarter |
| 8 | Patient ZIP code (5-digit), masked to 3-digit |
| 9 | Small country groups |
| 10 | Patient ZIP code (3-digit) |

**Synthetic regeneration of data.** As a result of the masking process detailed above, several attribute values of unique records are already suppressed in the provided dataset. In order to quantify the reidentification risk inherent to the original data, we synthetically regenerated plausible original values for the masked values by following the distribution of original values that were masked for each attribute, which is also given by the data provider. Figures S1 and S2 show such distributions for the "race" and "age" attributes, respectively. The distributions for all attributes are available in (*9*). We can see that the less frequent original attribute values are proportionally more frequently masked because, due to their rarity, they tend to yield unique records.

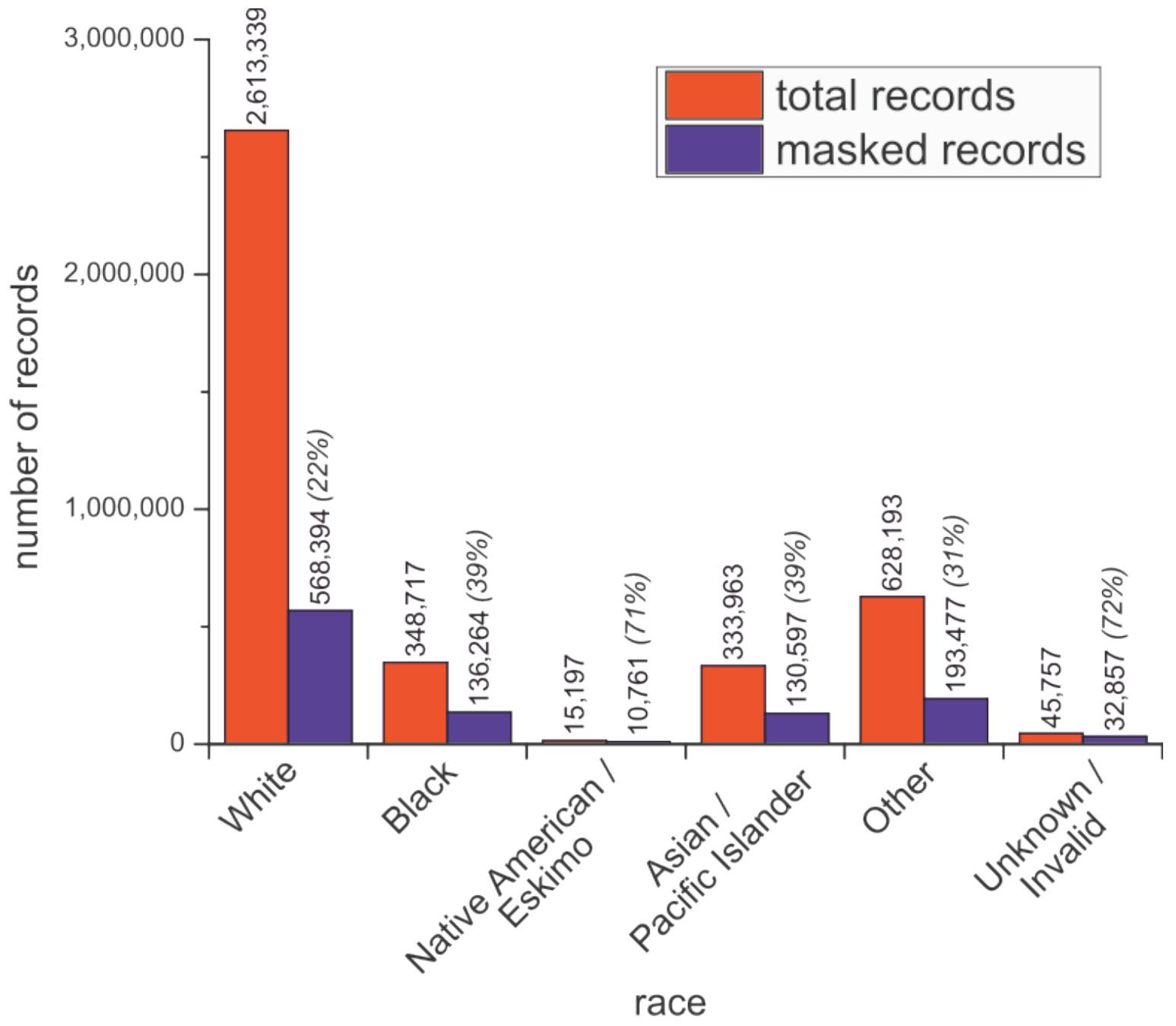

**Fig. S1. Total and masked record counts broken down by the "race" attribute**

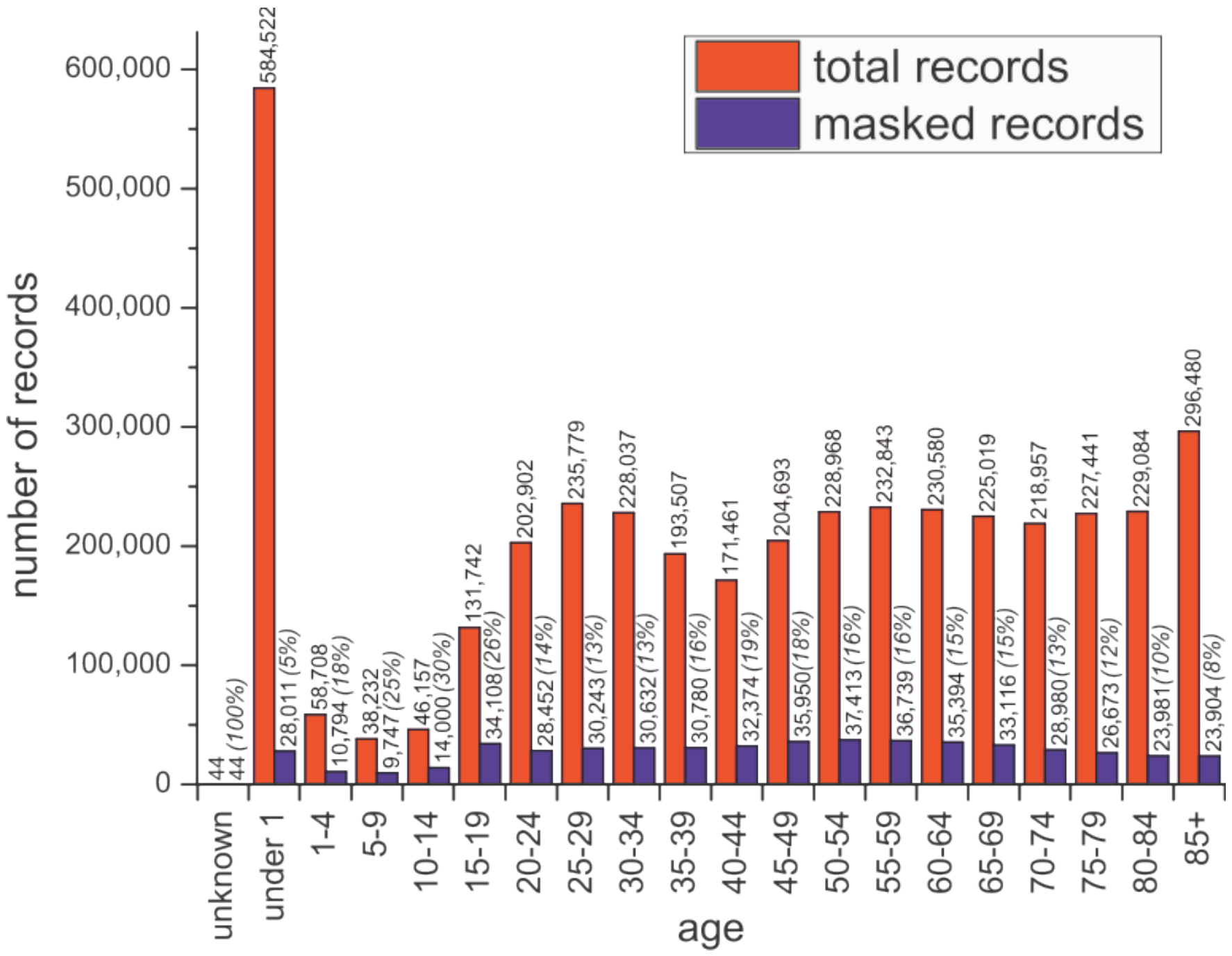

**Fig. S2. Total and masked record counts broken down by the "age" attribute**

According to these distributions of original attribute values that were masked, we created the partially synthetic dataset SPD by following Algorithm S1: for each attribute *a*, the algorithm takes every possible original value *e* and uses *e* to replace a number of the masked values for *a* equal to the absolute frequency of *e* in the distribution.

**Algorithm S1. Synthetic regeneration of masked values**

```
Inputs: D (PD dataset), matrix v[attribute][value] (with the distributions of original
values that were masked for all the attributes*)
Output: D^S (Partially synthetic dataset SPD)

D^S := D
for each attribute a in D^S do
   for each value e in a do
      R := randomly select a number v[a][e] of records in D^S with masked values
      for attribute a
      Assign value e to attribute a of records R in D^S
   end for
end for
output D^S
```

*For example, for the attribute "race" shown in Figure S1: *v*[race][1] = 568394, *v*[race][2] = 136264, *v*[race][3] = 10761, *v*[race][4] = 130597, *v*[race][5] = 193477, *v*[race][6] = 32857.

By the design of Algorithm S1, the SPD dataset exactly preserves the marginal attribute distributions of original data. To evaluate how well the SPD dataset preserves the correlations between attributes, we compared the pairwise correlations between the attributes in the de-identified (PD) and partially synthetic (SPD) datasets. Correlations in the PD dataset were measured from the attribute values that were not masked, which on average over the different attributes occur in 3,142,313 records. These correlations are reported in Table S4.

**Table S4. Pairwise correlations between quasi-identifying [unmasked] attribute values in the PD dataset**

| Attribute | age_yrs | sex | ethncty | race | patzip | patcnty | los | adm_qtr |
|---|---|---|---|---|---|---|---|---|
| **oshpd_id** | 0.001 | 0.001 | 0.045 | -0.056 | 0.043 | 0.811 | 0.003 | -0.001 |
| **age_yrs** | | 0.013 | 0.359 | -0.215 | -0.006 | -0.004 | 0.118 | -0.012 |
| **sex** | | | -0.048 | 0.022 | 0.007 | 0.012 | -0.036 | 0.004 |
| **ethncty** | | | | -0.334 | 0.134 | 0.036 | 0.048 | -0.007 |
| **race** | | | | | -0.081 | -0.054 | -0.014 | 0.015 |
| **patzip** | | | | | | 0.094 | -0.012 | -0.002 |
| **patcnty** | | | | | | | 0.001 | 0.000 |
| **los** | | | | | | | | -0.005 |

We can see that most attribute pairs are not correlated at all or show a very low correlation. Only the attribute pair Hospital ID (oshpd_id) and Patient County of Residence (patcnty) is highly correlated (0.811). This makes sense, since patients are normally admitted to a hospital near their residence (i.e., in the same county). In this case, if we decide to protect one of these attributes, we will also have to protect the other, since both disclose similar information about the patient (i.e., the hospital ID can be easily inferred from the county of residence and vice versa).

Table S5 shows the pairwise correlations after plausible original values have been synthetically generated for the masked values with Algorithm S1. In this case, the total 3,985,166 records (including attribute values that were not masked and values that were synthetically generated) have been considered.

**Table S5. Pairwise correlations for the partially synthetic quasi-identifying attributes**

| Attribute | age_yrs | sex | ethncty | Race | patzip | patcnty | los | adm_qtr |
|---|---|---|---|---|---|---|---|---|
| **oshpd_id** | 0.000 | 0.003 | 0.029 | -0.045 | 0.031 | 0.807 | 0.003 | -0.001 |
| **age_yrs** | | 0.006 | 0.209 | -0.117 | -0.001 | -0.004 | 0.022 | -0.006 |
| **sex** | | | -0.035 | 0.008 | 0.037 | 0.013 | -0.013 | 0.002 |
| **ethncty** | | | | -0.209 | 0.008 | 0.021 | 0.012 | -0.003 |
| **race** | | | | | -0.027 | -0.044 | 0.003 | 0.013 |
| **patzip** | | | | | | 0.180 | -0.014 | -0.007 |
| **patcnty** | | | | | | | 0.000 | -0.001 |
| **los** | | | | | | | | -0.002 |

Pairwise correlations in Tables S4 and S5 are very similar. In fact, the correlation between both sets of correlations is as high as 0.957. This shows that the partially synthetic dataset offers an accurate representation of the original data, not only at an attribute level (marginal distributions), but also at a record level (correlations). Such accuracy endorses the significance of the reidentification results we report in our study.

**Reidentification risk.** First, we must understand the kind of privacy threat at stake. Very often, databases contain *confidential* attributes about subjects together with personal identifiers (such as names or passport numbers) that uniquely identify the subject. Clearly, identifiers should be suppressed to avoid reidentification. However, often there are other attributes (such as zipcode, job or birthdate), called *quasi-identifiers* (QIs from now on), each of which does not uniquely identify the subject, but whose combination may. Since QIs may be present in public non-confidential databases (such as electoral rolls) together with some identifiers (such as a passport number), or may be known by a potential attacker on a target subject (e.g., a neighbor), they could be used to reidentify subjects in the confidential database, which would disclose the subjects' confidential attribute values. Thus, it is crucial to mask QIs before data release to prevent reidentification. It is easy to see that reidentification via QIs (treated in the literature on privacy protection at least since 1988 (*11*) and made popular in 1998 by the *k*-anonymity model (*12*)) is equivalent to the "unicity" idea re-discovered by dM in 2013 (*2*); that is, a subject whose quasi-identifying features are unique in a dataset risks being reidentified.

By following the above scenario, we assess the record reidentification risk according to the pieces of patient information (grouped as census, spatial and temporal features) assumed known to an attacker. For example, let us assume the attacker knows that his neighbor John Doe, who is a 42 year-old white non-Hispanic male living in 93722, Fresno County, was admitted to the Cedars-Sinai Medical Center. Thus, he knows two pieces of his neighbor's information (census details and admission location). If there is only one record in the dataset with that combination of attribute values (the first record in Table S2), the attacker can reidentify John and learn John's confidential data (hospital charges and diagnosis). This procedure is equivalent to the one used by dM to compute "unicities", but considering the whole population of patients, instead of a sample of 10,000 users as done by dM. Here we must note that *not even population uniqueness is really equivalent to reidentification: just knowing that some anonymized individuals have unique features in the population does not automatically yield their identities.* As discussed above, one needs to be able to link the anonymized records of the unique individuals to some external identified data source.

Specifically, we compute the reidentification risk in a dataset *D* for a set of quasi-identifying attributes *Q* as the ratio of unique records in the dataset:

$$reidentification\ risk(D,Q) = \frac{unique\ records(D,Q)}{total\ records(D)}. \qquad (1)$$

Unique records are those whose combination of quasi-identifying attribute values does not appear in any other record of the dataset. In a dataset with (partially) suppressed (replaced with the wildcard *) or generalized values (e.g., specific ages replaced by a range of years, 27→[20-30]), a record is counted as unique if there is no other record with identical attribute values or with value ranges that *fit* the former record. We next give two examples to illustrate what fitting means. For example, if there is a record with age=26 and there is another record with identical values for the rest of quasi-identifying attributes but whose "age" has been generalized to the range [24-28], the former will not be considered unique because knowing that the age of an individual is 26 does not lead to unequivocal record reidentification. On the other hand, if there is a record with an "age" generalized to the range [24-28], we will consider all possible age values in the range (24, 25, 26, 27 and 28) and we will look for other records in which each age value fits (that is, records having an identical age or an age range including the value); the former record will be considered unique if we are unable to find a fitting record for at least one of these values. This defines the worst-case reidentification criterion, because it assumes that the real value behind a generalized range is precisely the one for which there is no other fitting record in the dataset. If attributes are replaced with a wildcard (*, as done in the HIPAA-based masking), we consider that this wildcard covers the whole range of possible values. For example, if the age value is replaced by *, it would cover the whole range of ages, from 0 to 85; if a zip code has been masked as 937**, it will cover the range [93700-93799].

Figure S3 shows that the uniqueness/reidentification risk for the SPD dataset reaches almost 75% of patients (nearly 3 million) when the attacker knows the three sets of features of the patient. This percentage is coherent with the high "unicities" found in dM for their credit card dataset, and it is even more alarming because (i) our assessment truly quantifies population uniqueness/reidentification risk (even though in the very worst case) and (ii) knowing our QIs is more plausible in practice. Moreover, to achieve such high "unicities" dM needed to link four spatiotemporal tuples/records for each individual; this was possible because dM's dataset included masked identifiers in each record, so that all records corresponding to the same individual can be linked. In contrast, our results show that a large number of reidentifications can still happen even when preventing such linkage (allowing it is very rash in a sensitive data release, by the way). In comparison, *the reidentification risk of the de-identified PD dataset is much lower (0.26% in the worst case)*, thanks to some QI values being replaced by wildcards. Admittedly, if each record contained a masked identifier (as in dM's dataset), the reidentification risk would increase because the attacker would be able to link several admissions of the same patient. It is precisely to thwart such linkage-based record aggregation that the HIPAA safe harbor rules do not allow including identifiers in any form.

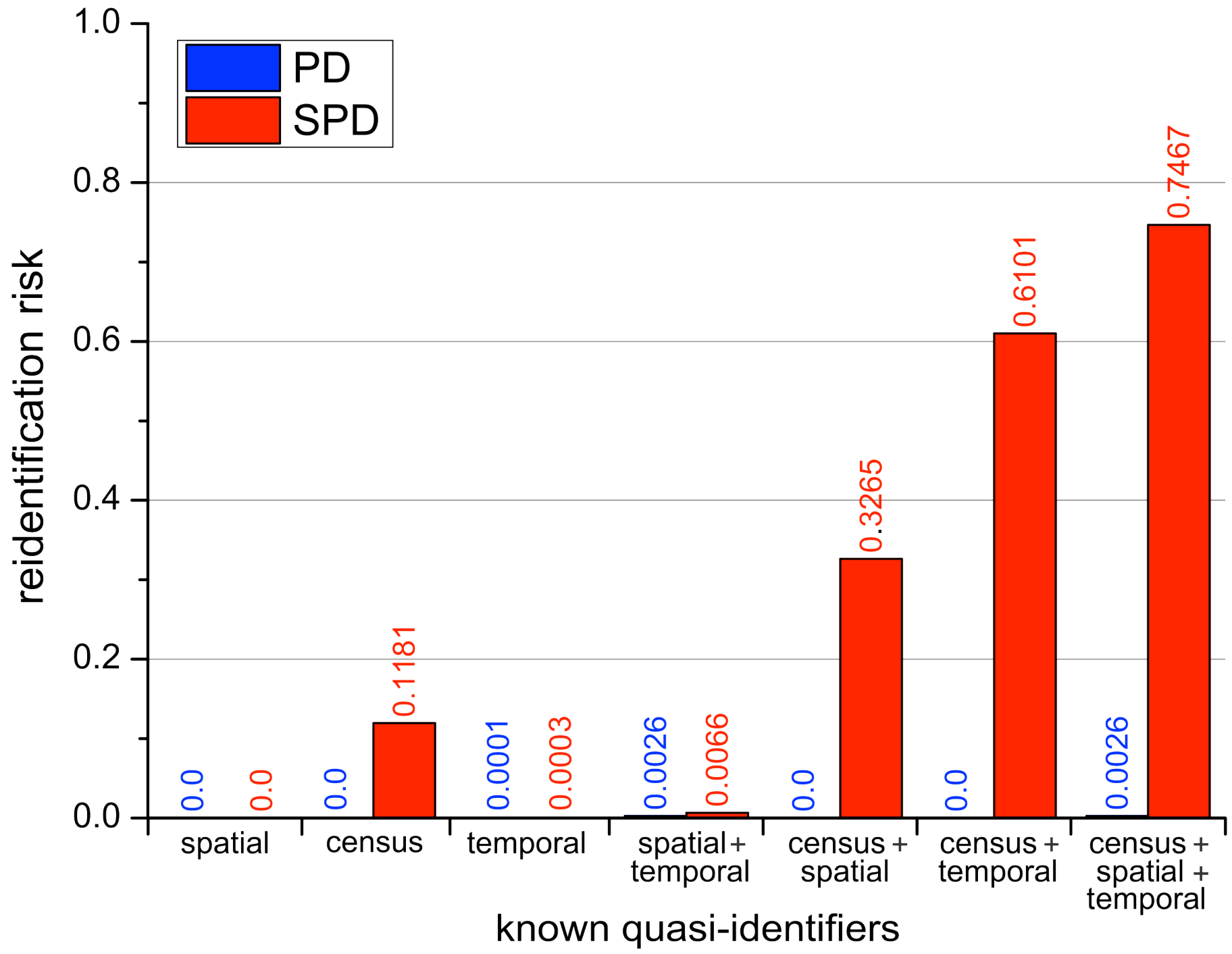

**Fig. S3. Reidentification risk in the PD and SPD datasets depending on the attributes known by the attacker**.

***k*-Anonymous data protection.** Reidentification has been treated in the *statistical disclosure control* (SDC) (*13, 14*) and *privacy-preserving data publishing* (PPDP) (*15*) literatures for nearly four decades. As a result, a broad choice of privacy protection mechanisms exists to balance the trade-off between protection (low disclosure risk) and utility needs (low information loss caused by anonymization) in the released data.

To illustrate the effectiveness of *sound* anonymization in front of the reidentification attack we are dealing with, we just need to resort to the well-known and simple *k*-anonymity notion (*12*). In a *k*-anonymous dataset, records should not include strict identifiers, and each one should be indistinguishable from, at least, *k*-1 other records, as far as the quasi-identifying attribute values are concerned. Thus, in a *k*-anonymous dataset the probability of reidentification of *any* individual is 1/*k*. Hence, for *k*>1, this probability is less than 1 for *all* records, thereby ensuring *zero unequivocal* reidentifications. Moreover, by tuning *k*, we can also tune the level of exposure of each individual. For example, in a 2-anonymous dataset, even though there are zero unequivocal reidentifications, there is still a 1/2 chance of *randomly guessing* the individual's record. If this is deemed too risky, we can increase the value of *k* (3 to 10 are usual choices (*16*)). Likewise, we can also lower the reidentification risk by considering additional attributes of the dataset as QIs and *k*-anonymizing all of them.

A method enforcing *k*-anonymity should detect records not satisfying the property and make their quasi-identifying values indistinguishable enough. Typical mechanisms to do this consist in grouping similar records in clusters of *k* or more, and *generalize/coarsen* their attribute values to their common range. In Table S6 we show a possible generalization-based *3*-anonymous version of the sample records shown in Table S2 for all QIs. Unlike dM's fixed and attribute-independent coarsening strategy (see details below), a *k*-anonymous method satisfies that:

- Records are grouped and generalized according to their similarity. This minimizes the information loss caused by value generalization (because the range of values in a group of similar records tends to be more compact) and, thus, preserves data utility better.
- The coarsening/generalization strategy is adapted to the data distribution at a *record level* so that (i) records *already* fulfilling *k*-anonymity *do not need to be modified* and, (ii) *all* the records that do not fulfill *k*-anonymity *will be* coarsened. In this manner, for *k*>1, we can *guarantee* that none of the records will be unequivocally reidentifiable while preserving the utility of the masked data as much as the privacy guarantee allows.

**Table S6**. **Example 3-anonymous version of the records in Table S2**

| *Quasi-identifiers* | | | | | | | | | *Confidential attributes* | |
|---|---|---|---|---|---|---|---|---|---|---|
| *Spatial feature* | *Census features* | | | | | | *Temporal features* | | | |
| **Hospital** | **Age** | **Gender** | **Ethnicity** | **Race** | **Zip code** | **County** | **Admission quarter** | **Length of stay** | **Charges** | **Diagnosis** |
| Cedars-Sinai | [42-52] | Male | Non-hispanic | [White, Black] | [93561-93722] | [Fresno, Kern] | [1-2] | [13-47] | $505,785 | Pneumonia |
| Cedars-Sinai | [42-52] | Male | Non-hispanic | [White, Black] | [93561-93722] | [Fresno, Kern] | [1-2] | [13-47] | $1,083,683 | Meningitis |
| Cedars Sinai | [42-52] | Male | Non-hipanic | [White, Black] | [93561-93722] | [Fresno, Kern] | [1-2] | [13-47] | $252,303 | Necrosis |
| Cedars Sinai | [57-70] | [Male, Female] | [Hispanic, Non-Hispanic] | White | [93551-93560] | [Kern, L.A.] | [2-4] | [1-2] | $107,179 | Breast Cancer |
| Cedars Sinai | [57-70] | [Male, Female] | [Hispanic, Non-Hispanic] | White | [93551-93560] | [Kern, L.A.] | [2-4] | [1-2] | $34,392 | Kidney Cancer |
| Cedars Sinai | [57-70] | [Male, Female] | [Hispanic, Non-Hispanic] | White | [93551-93560] | [Kern, L.A.] | [2-4] | [1-2] | $24,846 | Stomach Cancer |

Algorithm S2 details how we have implemented generalization-based *k*-anonymization. Given that all attributes in the PD dataset are numerical or are numerically coded, we sort them in ascending order, group them in clusters of *k* records and generalize the values of each attribute in the records of each cluster by replacing them with the range defined by the attribute's within-cluster minimum and maximum.

**Algorithm S2. Generalization-based *k*-anonymization**

```
Inputs: D (dataset), k (level of anonymity)
Output: D^A (a transformation of D that satisfies k-anonymity)

1   sorted_D := sort records in D in ascending order by the distance to a dummy
    reference record with all quasi-identifiers set to zero, where the distance is
    computed based only on the quasi-identifiers (Eq. 2)
2   D^A := sorted_D
3   while (|sorted_D| ≥ 2*k) do
4      create a cluster in D^A with the first k records in sorted_D
5      remove these records from sorted_D
6   end while
7   create a cluster in D^A with the remaining records in sorted_D
8   for each cluster c in D^A do
9      for each attribute a in D^A do
10         [min,max] := calculate the minimum and maximum values of attribute a over
           records of cluster c
11         Replace values of attribute a in records of cluster c by the range [min,max]
12     end for
13  end for
14  output D^A
```

**Distance between records.** Algorithm S2 sorts records according to their distance in terms of their quasi-identifiers. We calculate the distance between two records, $r_x$ and $r_y$, as the standardized Euclidean distance between quasi-identifying values, whereby the pairwise difference between attribute values is normalized by the standard deviation of the attribute (so that the aggregated distance is not biased towards the attributes with the largest value ranges). Mathematically,

$$distance(r_x, r_y) = \sqrt{\sum_{j=1}^{m} \left( \frac{(a_{xj} - a_{yj})}{\sigma_j} \right)^2}, \qquad (2)$$

where $a_{xj}$ and $a_{yj}$ are the values of the $j$-th quasi-identifying attribute for records $r_x$ and $r_y$, respectively, $\sigma_j$ is the standard deviation of the $j$th attribute in the dataset and $m$ is the number of quasi-identifying attributes.

**Naïve data coarsening implemented by dM to anonymize data.** dM correctly remarked that reidentification risk (or "unicities") greatly depend on the data resolution. Data coarsening is indeed a method to reduce data resolution and it is very often used in anonymization (e.g., to achieve *k*-anonymity (*12*), as described above). However, dM concluded that the coarsening they applied to anonymize data was not effective enough; as a result, they questioned the general effectiveness of anonymization methods. That their coarsening was ineffective is hardly surprising given its naïveness: attribute values were coarsened using value ranges that were fixed *a priori*; for example, times were replaced by time windows from 1 to 15 days and prices were replaced by fixed price intervals. This is clearly unsuitable for anonymization for at least two reasons: (i) coarsening should be based on the actual distribution of the dataset if true anonymization guarantees are to be offered (e.g., a fixed price range may contain a single price value among those in the dataset); (ii) independently coarsening each quasi-identifying attribute cannot guarantee that unique combinations of QI values disappear (coarsening must consider all QIs together).

Algorithm S3 details the (naïve) data coarsening method employed to reduce the resolution of attribute values that dM used in their study. The coarsening process is applied independently to each attribute by using a number (the target resolution) of fixed intervals that constitute a partition of the domain of the attribute. For example, for the domain [0-85+] of the "age" attribute, if we set the resolution to 4, the values will be coarsened into the intervals [0-20], [21-42], [43-63] and [64-85+]. Coarsenings of the "age" attribute for other resolutions are given in Tables S7 and S8.

**Algorithm S3. Data coarsening**

```
Inputs: D (dataset), r (vector with the target resolution for each attribute)
Output: D^A (coarsened D as per r)

D^A := D
for each attribute a in D^A do
   E :=[min(a),max(a)], where min(a) and max(a) are the smallest and greatest
   values in the domain of a, respectively
   partition E into r[a] subintervals each having approximately the same length
   for each subinterval S in E do
      [min,max] := calculate the minimum and maximum values of S
      replace values in the range [min,max] of attribute a in D^A by [min,max]
   end for
end for
output D^A
```

**Table S7. Intervals resulting for the "age" attribute when coarsened with resolution 16**

| # | Interval |
|---|---|
| 1 | [0-4] |
| 2 | [5-9] |
| 3 | [10-15] |
| 4 | [16-20] |
| 5 | [21-25] |
| 6 | [26-31] |
| 7 | [32-36] |
| 8 | [37-42] |
| 9 | [43-47] |
| 10 | [48-52] |
| 11 | [53-58] |
| 12 | [59-63] |
| 13 | [64-68] |
| 14 | [69-74] |
| 15 | [75-79] |
| 16 | [80-85+] |

**Table S8. Intervals resulting for the "age" attribute when coarsened with resolution 8**

| # | Interval |
|---|---|
| 1 | [0-9] |
| 2 | [10-20] |
| 3 | [21-31] |
| 4 | [32-42] |
| 5 | [43-52] |
| 6 | [53-63] |
| 7 | [64-74] |
| 8 | [75-85+] |

**Comparison between "anonymization" methods.** We empirically compared the risk of unequivocal reidentification and correct random reidentification in *k*-anonymity, dM's naïve coarsening (both applied to the SPD dataset) and the "safe harbor" de-identified PD dataset. For *k*-anonymity, we implemented the generalization-based mechanism introduced above (see Algorithm S2) and applied it to all QIs in SPD (spatial, census and/or temporal features) for several values of *k*. We also used dM's naïve coarsening on SDP, by performing fixed and attribute-independent coarsening of the different sets of attributes (Algorithm S3); specifically, we coarsened using intervals covering 1/32, 1/16 and 1/8 of the domain ranges of the attributes (e.g., Tables S7 and S8). For safe harbor, we took the [de-identified] PD dataset. Results are presented in Figure S4. We can see that *k*-anonymity yields zero unequivocal reidentifications and a rate 1/*k* of correct random reidentifications in the worst case, when the attacker knows the values of all QIs. In contrast, neither naïve coarsening nor safe harbor completely eliminate uniqueness and, hence, they do not eliminate unequivocal reidentifications.

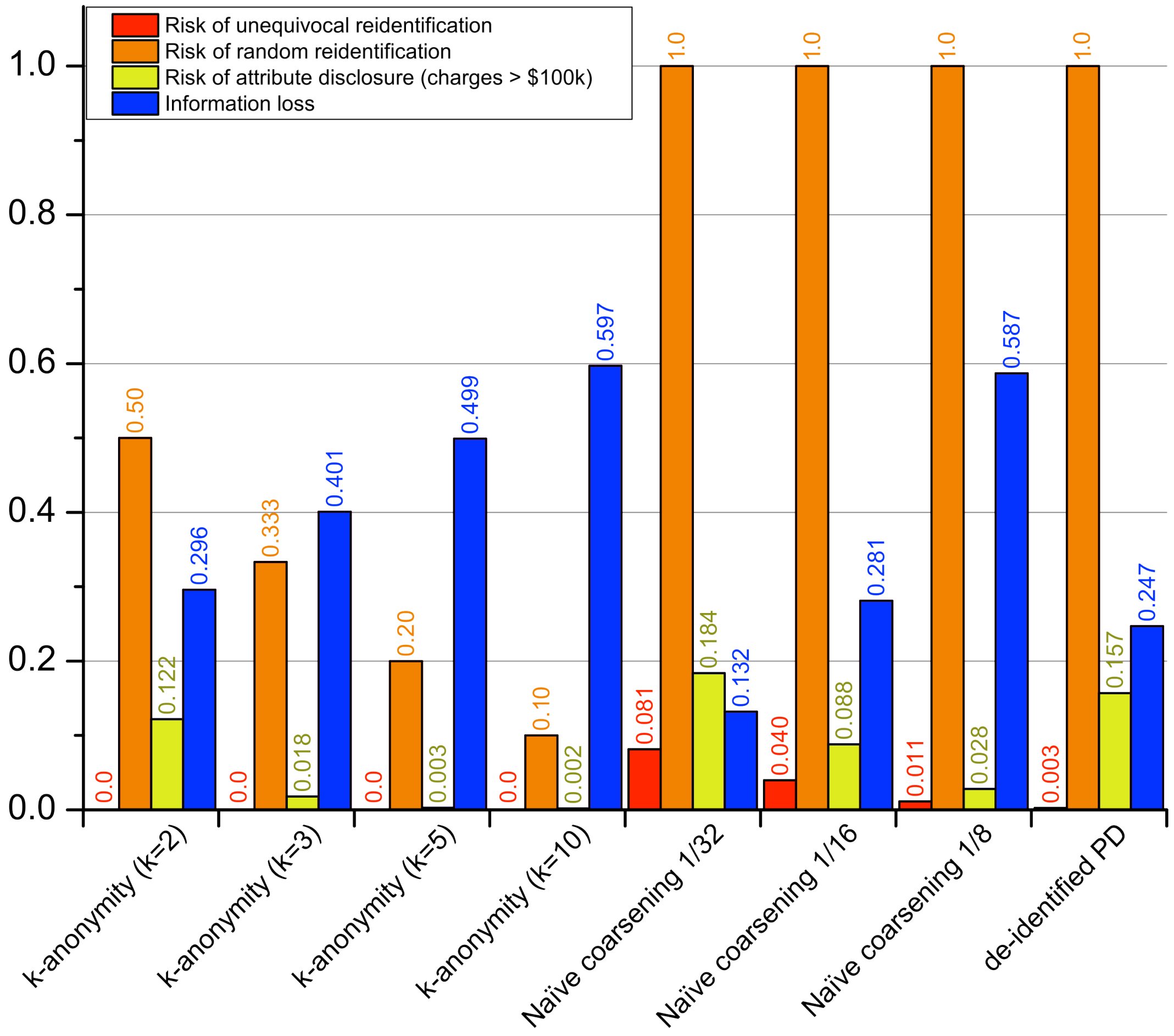

**Figure S4. Reidentification risk, attribute disclosure risk, and information loss for *k*-anonymized SPD, naïve coarsened SPD and "safe harbor" de-identified PD.**

To complement the above results on dM's data coarsening, Figure S5 provides a multidimensional view of the reidentification risk when the spatiotemporal and the census features are independently coarsened. We observe a sharp decline in the reidentification risk when coarsening to intervals covering 1/32 of the domain of the attributes, and an almost linear decrease as we duplicate the interval size. In any case, even using the most severe coarsening for all attributes (in two intervals covering each 1/2 of the attribute's domain, which causes a lot of information loss), there are still 51 records that can be unequivocally reidentified.

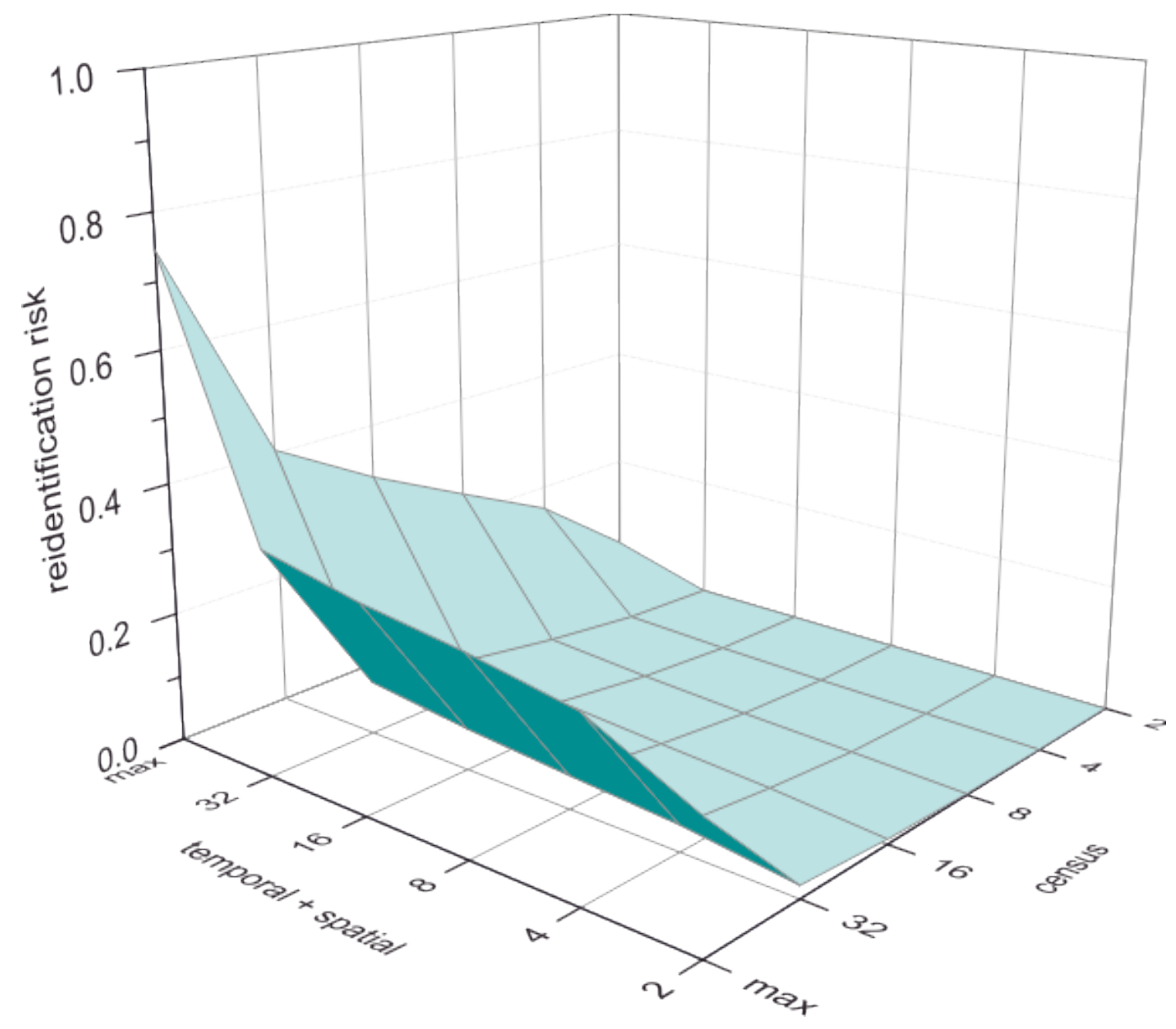

**Fig. S5. Reidentification risk when coarsening the spatiotemporal and census features of the SPD dataset into fixed intervals (between resolution 2 –two intervals covering the attribute range– and the maximum –original– resolution).**

**Data utility and information loss calculation.** The authors of dM only quantified the reidentification risk of the de-identified dataset. However, this is only half of the picture: the protected dataset should also remain analytically useful (which is the ultimate justification of data publishing). Note that, if we do not care about utility, encryption or even removal of the data are simpler and safer alternatives to anonymization. To illustrate this, we have also measured the utility of the outcomes of the different anonymization strategies: de-identified PD dataset, naïvely coarsened SPD dataset and SPD dataset *k*-anonymized over all QIs. Data utility preservation is usually quantified as the reciprocal of the *information loss* incurred by masking (*12*); that is, the less information loss, the more analytically useful the data remain. Information loss can be viewed as the distance between the original and masked values for all attributes and records. In a dataset *D'* masked with *k*-anonymity or data coarsening, the information loss with respect to the original dataset *D* can be calculated as

$$Information\ Loss(D, D') = \frac{1}{n \times m} \sum_{i=1}^{n} distance(r_i, r_i'), \qquad (3)$$

where $n$ is number of records, $m$ is the number of attributes and $r'_i$ is the masked version of record $r_i$. The distance between records is computed as per Eq. (2). However, since we are now comparing point values (i.e., original attribute values $a_{ij}$) with generalized/coarsened values ($a'_{ij}$, which correspond to the [*min*, *max*] intervals resulting from Algorithms S2 and S3), we transform $a'_{ij}$ into a point value that is the interval limit farthest from $a_{ij}$ (this is the worst-case information loss scenario). Mathematically, the transformation is

$$a'_{ij} = \begin{cases} \max\ of \left[\min, \max\right], & if \left(\max - a_{ij}\right) > \left(a_{ij} - \min\right); \\ \min\ of \left[\min, \max\right], & otherwise. \end{cases} \quad (4)$$

If masked values are replaced by wildcards, the maximum and minimum limits of the attribute's domain are considered.

In Figure S4 we see that *k*-anonymity not only outperforms naïve coarsening at yielding zero unequivocal reidentifications, but it does so with even less information loss (e.g., 2-anonymity vs. the safest naïve coarsening considered, that is, 1/8-naïve coarsening). Obviously, increasing *k* increases the information loss incurred by *k*-anonymity, but this is compensated by a decreasing reidentification probability (*1/k*). Naïve coarsening (see also Figures S5 and S6) cannot provide such privacy guarantee and trade-off, and is therefore a poor anonymization method.

Figure S6 complements Figure S4 by showing a multidimensional view of the information loss caused by naïve coarsening. We observe that coarsening the spatiotemporal features (which are the most fine-grained ones) causes a very large loss of information. Coarsening the census features, on the other hand, causes much less loss, since most of them are already expressed with a limited number of categories (e.g., "sex", "race" and "ethnicity" have each between 3 and 6 categories). Nonetheless, information loss is significantly greater than with *k*-anonymity in most cases (as reported in Figure S4), whereas the reidentification risk does not reach zero in any case.

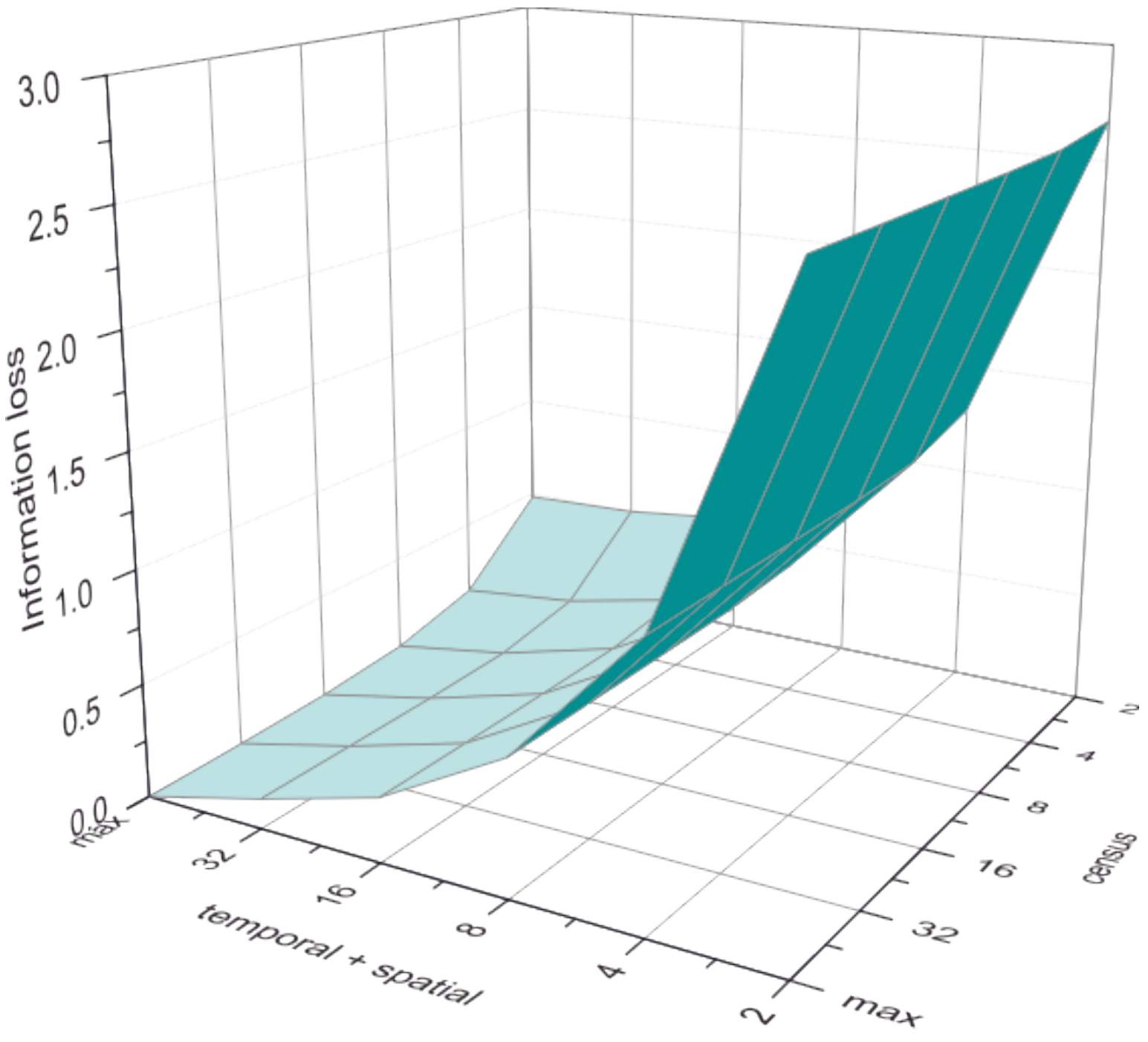

**Fig. S6. Information loss when coarsening the spatiotemporal and census features of the SPD dataset into fixed intervals (between resolution 2 –two intervals covering the attribute range– and the maximum –original– resolution).**

**Attribute disclosure risk.** While *k*-anonymity only protects against unequivocal reidentification, extensions of it address *attribute disclosure*, which occurs if the values of a confidential attribute within an equivalence class are too close. For example, in the 3-anonymous dataset shown in Table S6, we cannot unequivocally reidentify John Doe; at most, we can guess his record with probability 1/3. However, if we know that John's QIs match the generalized intervals of the first group of 3-anonymous records, we learn that his [confidential] hospital charges are very high (>$250,000), which may suggest that he is a wealthy person. Likewise, if we can establish that a patient's record belongs to the second group of 3-anonymous records, we learn that s/he suffers from cancer. We compute the disclosure risk of a confidential attribute value (e.g., diagnosis: AIDS positive) or value range (e.g., salary: above $150,000; diagnosis: *any* cancer type) *z* in an anonymized dataset *D'* as the proportion of records with value (range) *z* that belong to an equivalence class *C* in *D'* (i.e., a group of records with the same generalized/coarsened QIs) that is non-diverse, i.e. such that *all* other records in *C* also have value (range) *z*.

$$attribute\ disclosure\ risk(D',z) = \frac{\sum_{C \in non-diverse_equivalence_class(D',z)} |C|}{total\ records(D',z)}. \qquad (5)$$

To measure this risk, we defined *z* as the range of hospital charges above $100,000 (around 10% of the total) which may indicate particularly "wealthy" patients that can afford such high charges. Figure S4 shows that none of the methods considered so far is free from attribute disclosure, even though *k*-anonymity yields the best results. Indeed, for the other methods, the attribute disclosure risk is increased by their higher reidentification risk (i.e., if we can reidentify an individual we can unequivocally learn her confidential attributes).

***k*-Anonymity & *t*-closeness.** Models that protect against attribute disclosure include *l*-diversity (*17*), which requires at least *l* well-differentiated confidential attribute values in each group of indistinguishable records, and *t*-closeness (*18*), which requires the distribution of each confidential attribute within each group to be similar to the distribution of the attribute in the entire dataset. To show its effectiveness, we added *t*-closeness to *k*-anonymity by using Algorithm S4 (adapted from (*19*)). For the sake of clarity in the following sketch of the algorithm, we make the simplifying assumption that *k* divides the number *n* of records (see (*19*) and Algorithm S4 for the general case). We first sort the records in the dataset in ascending order of the confidential attribute and split them into *k* subsets of consecutive (*n*/*k*) records. Then, to mimic the distribution of the confidential attribute over the entire original dataset within each *k*-anonymous group, we create equivalence classes with *k* records by iteratively picking one record from each of the subsets and, finally, generalizing the QIs of each resulting group of *k* records to a common range. Because the subsets have been created after sorting by the confidential attribute, this process will yield equivalence classes with uniformly distributed samples of the confidential attribute, which is the goal of *t*-closeness. To maximize the homogeneity of the original QI values of the records assigned to each equivalence class and, thus, minimize the information loss resulting from their generalization, we also sort records within each subset of (*n*/*k*) records in ascending order of the quasi-identifier attributes.

Even though, given a value of *k*, there is freedom to choose the value of the *t* parameter, by the design of our algorithm, *t* is upper-bounded as follows for a dataset with *n* records:

$$t \le \frac{(n-k)}{(2(n-1)k)}. \tag{6}$$

Conversely, for a given value of *t*, the minimal cardinality *k'* of the clusters for a dataset with *n* records is:

$$k' \ge \frac{n}{2(n-1)t+1}. \tag{7}$$

Thus, the actual cluster size (i.e., actual *k*-anonymity level) that the algorithm will produce to fulfill the desired *k*-anonymity and *t*-closeness levels is:

$$\mathit{cluster\ size} = \max\left\{k, \left[\frac{n}{2(n-1)t+1}\right]\right\}. \tag{8}$$

**Algorithm S4. Generalization-based *k*-anonymization & *t*-closeness**

```
Inputs: D (dataset), k (level of anonymity), t (level of closeness)
Output: D^A (a transformation of D that satisfies k-anonymity & t-closeness)

k = max[k, n/(2(n-1)t+1)]
k = k+[(n mod k)/(n/k)]
sorted_D := sort records in D in ascending order of the confidential attribute
split sorted_D into S_1,..,S_k subsets with (n/k) records, create a final subset with
the remaining (n mod k) records
sort records in each subset S_i in ascending order by the distance to a dummy
reference record with all quasi-identifiers set to zero, where the distance is
computed based only on the quasi-identifiers (Eq. 2)
while (|sorted_D| > 0) do
   cluster C = Ø
   for each subset S_i in sorted_D do
      add to cluster C the first record in S_i and remove it from sorted_D
   end for
   for each attribute a in C do
      [min,max] := calculate the minimum and maximum values of attribute a over
      records of cluster C
      replace values of attribute a in records of cluster C by the range [min,max]
   end for
   add to D^A all records in cluster C
end while
output D^A
```

In all the experiments described in this study, we set the *t*-closeness level to the upper bound given by the desired level of *k*-anonymity (according to Eq. (6)). Table S9 specifies the value of *t* for each *k*.

**Table S9. *t*-Closeness values corresponding to the desired levels of *k*-anonymity for the SPD dataset (*n*=3,985,166 records)**

| *k* | *t* |
|---|---|
| 2 | 0.25 |
| 3 | 0.17 |
| 5 | 0.10 |
| 10 | 0.05 |

Figure S7 compares *k*-anonymity with and without *t*-closeness, both regarding attribute disclosure risk (lowered by *t*-closeness to zero in all cases) and information loss (slightly increased by *t*-closeness as a result of grouping records with less homogeneous QIs, but in most cases less than the loss caused by the safest naïve coarsening considered in Figure S4, that is, 1/8-naïve coarsening).

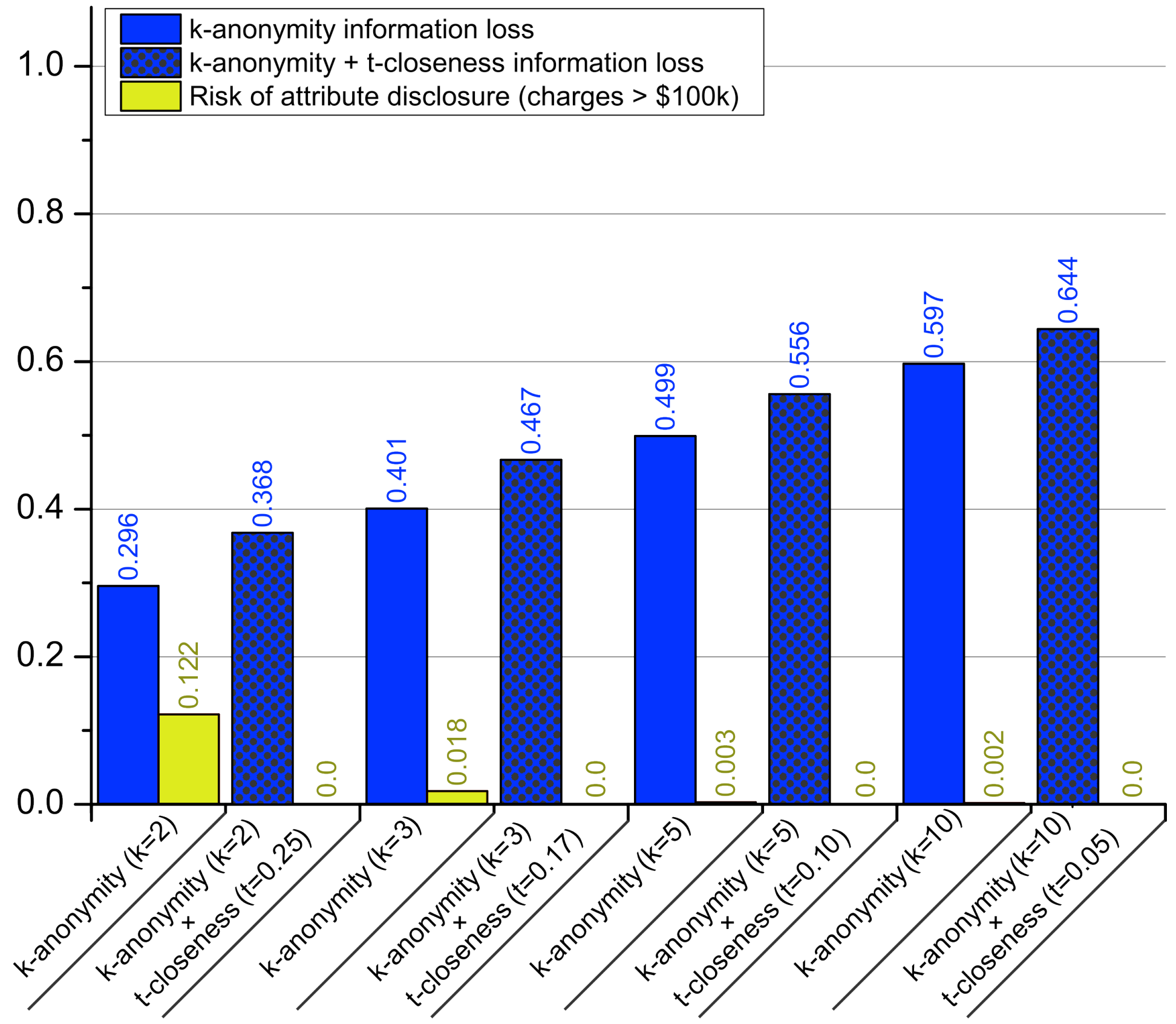

**Fig. S7. Attribute disclosure risk (for charges above $100,000) and information loss for *t*-closeness & *k*-anonymity vs. plain *k*-anonymity**

**Other data anonymization mechanisms**. So far, we have only resorted to generalization-based *k*-anonymity and *t*-closeness to protect structured datasets. Yet, there is much more in the anonymization literature.

First, we can find *k*-anonymity extensions designed for other dataset types. For datasets in which individual transactions should still be linkable, $k^m$-anonymity requires that each combination of up to *m* transactions (where *m* is the maximum number of transactions assumed known by an attacker) must appear at least *k* times in the anonymized dataset (*20*). For collections of unstructured documents (like e-mails), *K*-safety states that the textual contexts of the entities to be protected (e.g., names, sensitive diseases) should be made indistinguishable in groups of, at least, *K* documents (*21*).

Second, the masking methods available to satisfy the above privacy models are not limited to coarsening/generalization. Alternatives include perturbative methods (noise addition, microaggregation, permutation, etc., see (*14*)), which have the advantage of preserving the granularity and the variability of the original data. This in turn enhances the utility of the masked data, because a range of statistics can be preserved.

Finally, beyond the classical data release anonymization considered in this study, the current research agenda includes more challenging scenarios, like anonymizing answers to interactive queries (for example, using ε-differential privacy (*22*), a privacy model also usable to anonymize data releases (*23*)), *big data* anonymization (in which scalability and some linkability preservation between heterogeneous anonymized sources are crucial) (*24, 25*), streaming data anonymization (*26*) and anonymization by the data subjects themselves (be it local anonymization (*27*) or co-utile collaborative anonymization *(28)*).

**Data availability.** The public de-identified PD dataset can be obtained on request at the source: http://www.oshpd.ca.gov/HID/Products/PatDischargeData/PublicDataSet/. To reproduce the experiments in our study, the zipfile available at http://crises-deim.urv.cat/opendata/SPD_Science.zip includes the partially synthetic quasi-identifiers (which correspond to the SPD dataset) and their *k*-anonymous, *k*-anonymous & *t*-close, and coarsened versions (with the confidential attribute *hospital charges*). Table S10 details the datasets that can be found in the zipfile.

**Table S10. List of dataset files provided for replication**

| File name | Description |
|---|---|
| SPD_dataset.txt | Partially synthetic quasi-identifiers generated with Algorithm S1 |
| 2_anonymous_SPD.txt | 2-anonymous version of the SPD dataset, with *charges* |
| 3_anonymous_SPD.txt | 3-anonymous version of the SPD dataset, with *charges* |
| 5_anonymous_SPD.txt | 5-anonymous version of the SPD dataset, with *charges* |
| 10_anonymous_SPD.txt | 10-anonymous version of the SPD dataset, with *charges* |
| 2_an_025_clos_SPD.txt | 2-anonymous and 0.25-close version of SPD, with *charges* |
| 3_an_017_clos_SPD.txt | 3-anonymous and 0.17-close version of SPD, with *charges* |
| 5_an_010_clos_SPD.txt | 5-anonymous and 0.10-close version of SPD, with *charges* |
| 10_an_005_clos_SPD.txt | 10-anonymous and 0.05-close version of SPD, with *charges* |
| 32_coarsened_SPD.txt | Coarsened version of the SPD dataset in intervals covering each 1/32 of the value ranges of all the attributes, with *charges* |
| 16_coarsened_SPD.txt | Coarsened version of the SPD dataset in intervals covering each 1/16 of the value ranges of all the attributes, with *charges* |
| 8_coarsened_SPD.txt | Coarsened version of the SPD dataset in intervals covering each 1/8 of the value ranges of all the attributes, with *charges* |

All datasets are in plain text format. The first line specifies the names of the attributes. Each successive line contains the comma-delimited [original/regenerated/masked] quasi-identifying attribute values of each record. If appropriate, the confidential attribute is included at the end of the line. When attribute values are generalized/coarsened to an interval, the *min* and *max* bounds of the interval are enclosed with brackets and delimited with a semicolon. All data files present records in the same order as in the original PD dataset so that they can be directly compared. With these data files (plus the original PD dataset obtained from the source), Figures S1-S3 and Tables S4 and S5 can be recreated. Also, by applying Algorithms S2, S3 and S4 to the partially synthetic SPD dataset, the *k*-anonymous, *k*-anonymous & *t*-close, and coarsened versions of the dataset and Figures S4-S7 can be recreated.

**Source code**. The code available at http://crises-deim.urv.cat/opendata/SPD_Science.zip offers an open source implementation of the three anonymization algorithms detailed above (Algorithm S2: *k*-anonymity; Algorithm S3: naïve coarsening; Algorithm S4: *k*-anonymity & *t*-closeness). Even though the code is specifically tailored to the synthetically generated SPD dataset we used in our experiments, it can also be applied to any structured dataset provided that the input data are in the same format.

As stated above, the input data consist of a text file that describes records as a list of attribute values. Each line corresponds to a record except the first line, which is the header and contains the names of attributes, delimited by commas. The last attribute is assumed to be the confidential attribute, whereas the preceding ones (no matter their number) are assumed to be the quasi-identifiers. For example, the header for the synthetic-generated SPD dataset is:

```
oshpd_id,age_yrs,sex,ethncty,race,patzip,patcnty,los,adm_qtr,charge
```

In this case, the last attribute "charge" is the confidential one.

The rest of the lines in the file contain the attribute values of the records, delimited by commas. The attribute values must appear in the same order as the corresponding attribute names in the header line. An example record in the SPD dataset is:

```
380929,25,1,2,1,94928,49,18,4,220449
```

The source code has been written in Java (version 7). The "functions.java" class contains the implementation of the anonymization algorithms detailed above. The main class "test.java" contains several usage examples of those algorithms. When executed, the program generates an anonymized version of the input data and the console outputs the quality metrics mentioned in the report: the information loss incurred by the anonymization process, the number of unique records that still remain and the number of records that may result in attribute disclosure for the confidential attribute.

To run the algorithms on a large dataset (such as the SPD dataset, which contains almost 4 million records), it is recommended to assign additional memory to the Java Virtual Machine. To do so, once the source code has been compiled, the tests can be executed with the following command, which sets the size of the JVM's heap to 7GB:

```
>java -Xms7144m -Xmx7144m urv.crises.anonym.Test
```

## Summary.

We have illustrated that, if sound privacy models and methods proposed in the literature over the last decades are appropriately used, anonymization can effectively protect against reidentification and attribute disclosure while preserving substantial data utility. Thus, we hope to have shown that there is no need to reinvent well-known privacy notions ("unicity") and that the ineffectiveness of anonymization claimed by dM is due to the choice of a poor anonymization method (fixed and attribute-independent coarsening) and to a general disregard of 40 years of literature (privacy models, anonymization methods, metrics, etc.). While the concerns spread by dM's conclusions can only be explained by a lack of awareness of the relevant literature, they have the potential of very seriously undermining the willingness of data subjects and data owners to share data for research. Therefore, it is imperative to disprove such conclusions at the technical level, which was the main purpose of our study.

Furthermore, we also wanted to convey to a broad audience that sound privacy models and anonymization methods have been developed over the last decades. This background makes it possible to produce anonymized data with intuitive and robust privacy guarantees against reidentification, while retaining sufficient analytical utility. In our view, this ought to reassure data subjects, data owners and data users and restore their trust in de-identified/anonymized data sharing.

## References.